\documentclass[preprint,12pt]{elsarticle}

\usepackage{graphics}
 \usepackage{graphicx}
\usepackage{epsfig}
\usepackage{subcaption}
\usepackage{caption}
\usepackage{amssymb}
\usepackage{amsthm}
\usepackage{amsmath}
\usepackage{lineno}
\usepackage[margin=1in]{geometry}
\usepackage{multirow}
\journal{TSWJ}
\begin{document}

\begin{frontmatter}


\title{Mobility-Assisted On-Demand Routing Algorithm for MANETs in the Presence of Location Errors\tnoteref{t1},\tnoteref{t2},}
\tnotetext[t1]{This work was supported by the 2013 Research Fund of University of Ulsan.}
\tnotetext[t1]{Manuscript received 7 March 2014; revised and accepted 30 April 2014; published 18 May 2014. Academic Editor: Zhongmei Zhou.}
\author{Trung~Kien~Vu}
\ead{trungkien@mail.ulsan.ac.kr}
\author{Sungoh~Kwon\corref{cor1}}
\ead{sungoh@ulsan.ac.kr}
\cortext[cor1]{Corresponding author}
\address{School of Electrical Engineering \\University of Ulsan \\Ulsan, Korea, 680-749}

\begin{abstract}
In this paper, we propose a mobility-assisted on-demand routing algorithm for mobile ad-hoc networks in the presence of location errors. Location awareness enables mobile nodes to predict their mobility and enhances routing performance by estimating link duration and selecting reliable routes. However, measured locations intrinsically include errors in measurement. Such errors degrade mobility prediction and have been ignored in previous work. To mitigate the impact of location errors on routing, we propose an on-demand routing algorithm taking into account location errors. To that end, we adopt the Kalman filter to estimate accurate locations and consider route confidence in discovering routes. Via simulations, we compare our algorithm and previous algorithms in various environments. Our proposed mobility prediction is robust to the location errors.
\end{abstract}


\end{frontmatter}


\section{Introduction}
\label{Introduction}
A Mobile Ad-Hoc Networks (MANET)~\cite{Ismail07a} consists of a set of wireless mobile nodes that dynamically exchange data among themselves without relying on any fixed infrastructure. Because of their easy deployment and extension, MANET application scenarios include emergency and rescue operations, conference settings, car networks, personal networking, and so on. Due to limited transmission ranges and infrastructure-free networks, each node in such networks has the responsibility not only to discover new routes but also to relay messages.

The most challengeable problem of MANETs~\cite{Corson99a} is how to adapt the topology changing that affects the performance of the network~\cite{Camp02a, Bai03a}. Due to changeable topology, routes from sources to destinations may be suddenly broken and nodes have to discover other available routes to deliver data. The ad-hoc on-demand distance vector routing algorithm (AODV) was proposed as a reactive routing algorithm to allow mobile nodes to quickly adapt to topology changes and link breaks in mobile ad-hoc networks~\cite{P99ad}. To find a possible route, the AODV makes a source flood a routing request message over the network and discovers a route based on the principle of the shortest path. The amount of overhead messages for route discovery and route maintenance depends on the longevity of routing paths. The awareness of link and path durations can improve routing performance in such mobile networks~\cite{Tsao06, Jiang05, Luo10}.

In~\cite{Fan04h, Namuduri12}, the authors modeled the distribution of path duration and analyzed the relation between path duration and other factors such as relative speed, transmission range, and number of hops. Their analysis shows that routing protocol with higher path duration can improve the network performance. In~\cite{La07}, the authors also investigate the distribution of path duration and then design a scheme to select a route with the largest expected duration and provide reliable network services in MANETs.

Location information enables nodes to predict mobility and estimate path durations more accurately. In~\cite{su2001, HU11, Chao07}, the authors proposed schemes to improve routing performance with location awareness. The proposed algorithms in~\cite{su2001, HU11} anticipate the link expiration time~(LET) based on measured locations and velocities, and apply for routing protocols to reduce overheads in~\cite{su2001} or to select the most reliable route that has the longest path duration~\cite{HU11}. In~\cite{Chao07}, the link duration time is adaptively applied for route maintenance in order to reduce unnecessary overhead. However, lifetime of link may be incorrectly calculated due to location errors that lead to incorrect hello frequency setting.

In practical deployment scenarios, location errors intrinsically occur in measurement~\cite{skwon06a}, even if locations are measured by the global positioning system~(GPS) receiver. Such imperfect location information leads imperfect mobility prediction, which results in performance degradation. However, the previous work assumed error-free location information and developed routing algorithms. In~\cite{su2001}, the impact of location errors on routing performance was provided only by simulations, but there is no effort to improve routing performance in such noisy information environments. Therefore, it is necessary to develop an efficient routing that is robust to location errors.

In this paper, we proposed a mobility-assisted on-demand routing algorithm in the presence of location errors in order to mitigate the impact of location errors on routing performance. To that end, the algorithm adopts the Kalman filter to compensate for the measurement location errors and estimates link durations to reduce overheads and select reliable routes. We also consider the confidence level of route in selecting the best route. Via simulations, we compare our proposed algorithm with previous algorithms.

The rest of this paper is organized as follows. In Section~\ref{Sys-Model}, we describe the system model and problem. In Section~\ref{ProposedAlgorithm}, we propose a Kalman filter based routing algorithm with mobility prediction for location correction and route selection. In Section~\ref{Evaluation}, we provide numerical results to analyze the impact of location errors and the efficient of our proposal in the presence of location errors, and we conclude the paper in Section~\ref{Conclusion}.

\section{System Model and Problem}
\label{Sys-Model}
\subsection{System model}
\label{Assumptions}
In this paper, we consider a mobile wireless network that supports multi-hop routing. The network is modeled as a set $\mathcal{N}$ of mobile nodes with transmission range~$r$ and a set $\mathcal{L}$ of communication links~$(i,j)$ between nodes~$i$ and $j$ in $\mathcal{N}$.

Link~$(i,j)$ is called \emph{valid} or \emph{connected} link at time~$t_k$ when the distance between nodes~$i$ and $j$ at time~$t_k$ is less than or equal to the transmission range~$r$, i.e., \[|X_{i}(t_k) - X_{j}(t_k)| \leq r,\]
where $X_{i}(t_k)$ and $X_{j}(t_k)$ are locations of nodes~$i$ and $j$, respectively, and $|X|$ stands for a Euclidian distance of vector $X$. Otherwise, link$~(i,j)$ is considered \emph{broken} or \emph{disconnected}, because the two nodes are out of their communication range. The link duration of link~$(i,j)$ is defined as the time interval for which the link is valid.

Due to a limited transmission range, packets are delivered from a source to a destination in a multi-hop manner via a route, which is defined as a set of links. For given source and destination nodes, $s$ and $d$, respectively, $H$ possible routes at time $t_k$ are denoted as $R_{(s,d)}^{(h)}(t_k)$ for $h \in \mathcal{H}=\{1, \cdots, H\}$, which consists of $\left|R_{(s,d)}^{(h)}(t_k) \right|$ links.

To find a route from a source to a destination and maintain routes, each mobile node employs the AODV routing algorithm, which is one of reactive routing protocols and frequently adopted in Mobile Ad-hoc Networks.

\subsection{Overview of AODV}
\label{Routing-Protocol}
The AODV~\cite{P99ad} routing algorithm consists of two main operations: route discovery and route maintenance. Route discovery is initiated by a source node that has data to send a destination node and does not have an active route in its routing table. To find a valid route to the destination, the source node broadcasts a route request~(RREQ) message, including a sequence number, to neighboring nodes. The RREQ message is flooded through the entire network until the message reaches the destination or an intermediate node that has a valid route to the destination. Each node that receives the RREQ message stores a reverse route to the source and then broadcasts the message to their neighboring nodes if the node is not the destination and the RREQ message is not a duplicate. When the RREQ message arrives at a destination node or an intermediate node that has a valid route to the destination, that the node sends a route reply~(RREP) message to the neighboring node in a reverse route in a unicast manner. The RREP message contains the number of hops to reach the destination node and the sequence number for the destination. A node receiving the RREP message sends this message to the source via the stored reverse route and then creates or updates a forward route to the destination.

Route maintenance is performed by nodes after route discovery operation, in order to maintain local connectivity and routes. Nodes periodically send a hello message to their neighbors to check if links are connected. If a node does not receive any hello message from its neighbors during a certain time period, referred to as the lifetime of hello message, the node assumes that the link to the neighbor is currently disconnected and reports the link failure to the source corresponding to the link via a route error~(RRER) message.

\subsection{Location Awareness and Performance Enhancement}
\label{LocationAwareness}
In a mobile ad-hoc network, the location information of nodes helps to improve routing performance, such as packet delivery rate and overhead by estimating node mobility. In a route discovery operation, the route with the longest lifetime can be selected to reduce the number of transmission failures and the number of overheads to find a new route~\cite{HU11}. To reduce overhead messages, instead of a fixed period for hello message, the adaptive period is proposed using link lifetime to achieve high protocol efficiency in~\cite{Chao07}.

To predict mobility, the previous work proposed a location prediction scheme~\cite{su2001}, which is defined as
\begin{eqnarray}\label{eq:futurelocation}
   \hat{X}_{i}(t_k + \Delta t) &=& X_{i}^{'}(t_k) + \overrightarrow{V}_{i}(t_k) \Delta t, 
\end{eqnarray}
where $\hat{X}_{i}(t_k +\Delta t)$, $X_{i}^{'}(t_k)$, and $\overrightarrow{V}_{i}(t_k)$ are the predicted location of node~$i$ at time $t_k + \Delta t$, a measured location at time~$t_k$, and a measured velocity at time~$t_k$, respectively. If individual velocities of nodes are not available in~(\ref{eq:futurelocation}), the nodes can approximately estimate their velocities using the previously stored location information~\cite{skwon06a} as follows: for $t_k > t_{k-1}$, the velocity of node~$i$ at time $t_k$ is approximately expressed as
\begin{eqnarray}\label{eq:futurespeed}
\overrightarrow{V}_{i}(t_k) \simeq \frac{X_{i}^{'}(t_k) - X_{i}^{'}(t_{k-1})}{t_k - t_{k-1}}.
\end{eqnarray}

Based on the mobility prediction, nodes estimate link durations corresponding to adjacent nodes, and destination nodes choose the longest lifetime route among candidates. Since a link between two nodes is connected only if the distance between the two nodes is less than or equal to their transmission range, the estimated link duration $LDT_{(i,j)}$ between nodes~$i$ and~$j$ is defined as
\begin{eqnarray}\label{eq-ldt}
  &LDT_{(i,j)} = & \max \Delta t \nonumber \\
  &\mathrm{subject \ to} & \hat{D}_{(i,j)}(t_k + \Delta t) \leq r,
\end{eqnarray}
where $\hat{D}_{(i,j)}(t_k + \Delta t)$ is the estimated distance between nodes~$i$ and $j$ elapsed time $\Delta t$ from current time~$t_k$. A route consists of ordered links, and is disconnected if one of the links is broken. Hence, the route expiration time~$RET_{(s,d)}^{(h)}$ of a route~$R_{(s,d)}^{(h)}$ between nodes~$s$ and $d$ is expressed as
\begin{eqnarray}\label{eq-ret}
  RET_{(s,d)}^{(h)} = \min_{(i,j) \in R_{(s,d)}^{(h)}} LDT_{(i,j)}.
\end{eqnarray}
for $h \in \mathcal{H}$. The most reliable route can be chosen among candidate routes based on~(\ref{eq-ret}).

\subsection{Location Errors and Estimation Problem}
\label{LocationErrors}
In practice, location errors inevitably exist in measurement. However, in previous work, mobility prediction used perfect location information receiving from the GPS devices or other techniques~\cite{misra2011, Drawil13}. The imperfect location information induces erroneous mobility estimate, which results in performance degradation.

\begin{figure}
  \centering
  \includegraphics[width=4.0in]{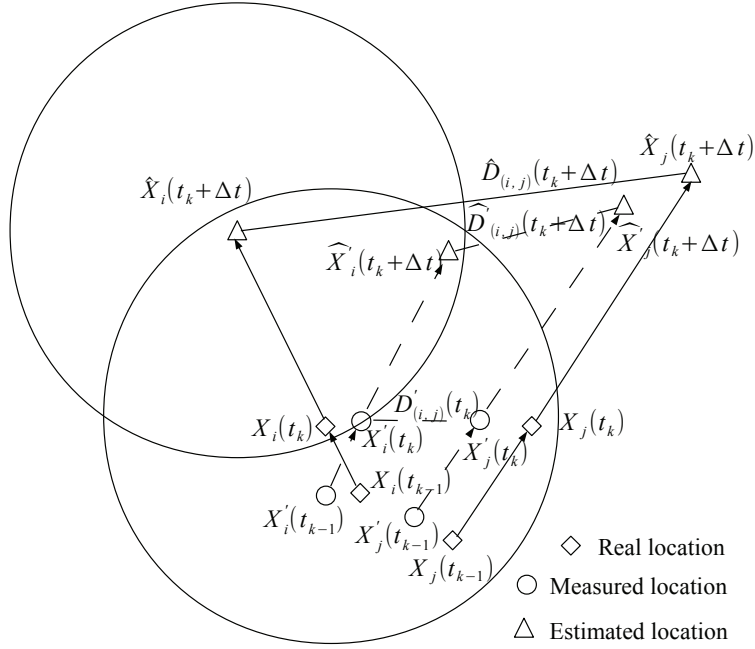}
  \caption{Estimated link duration}\label{System}
\end{figure}

For example, let $X_{i}(t_k)$ and $X_{i}^{'}(t_k)$ be the real location and the measured location of node~$i$ at time~$t_k$. Then, based on measured locations $X_{i}^{'}(t_k)$ and $X_{j}^{'}(t_k)$ of nodes $i$ and $j$, respectively, after elapsed time~$\Delta t$ from time~$t_k$, the estimated distance $\widehat{D}_{i}^{'}(t_k + \Delta t)$ between the two nodes is less than the transmission range~$r$ and the link between two nodes is considered \emph{connected}, even though node $j$ locates out of the transmission range of node~$i$, i.e, the communication link between two nodes is \emph{disconnected}, as shown in Fig.~\ref{System}. Hence, we propose a routing algorithm in the presence of location errors in measurement to mitigate the impact of imperfect location information.

\section{Proposed Algorithm}
\label{ProposedAlgorithm}
In this section, we proposed an on-demand routing algorithm robust to location errors with mobility prediction. In MANETs, the mobility prediction plays a great role in predicting the link lifetime and the route lifetime, which can reduce overhead messages and improve routing performance~\cite{HU11}. However, as shown in Fig.~\ref{System}, location errors in measurement provide an incorrect mobility prediction, which induces wrong decision for routing. To mitigate the impact of such errors on mobility prediction and routing decision, we adopt two schemes: location error correction and route confidence.

\subsection{Location Correction and Mobility Prediction}
\label{KalmanFilter}
We employ the discrete Kalman filter, which is a set of recursive mathematical equations and supports the estimation of states in such way that minimizes the variance of estimation errors. The recent updates with previous measured location compensates current location for measurement errors. In this paper, the process errors are ignored, the main focus is the measurement errors. A detail of the discrete Kalman filter is presented in~\cite{greg2006}.

From~(\ref{eq:futurelocation}), the current or future location depends on the previous location. The location errors is defined as the difference between the actual location and the measurement location. Let $W_{i}$ be the location errors at node $i$, which is the additive noise, then the measurement location of node $i$ at time $t_k$ can be expressed as $X_{i}^{'}(t_k) = X_{i}(t_k) + W_{i}(t_k)$.

For each node $i \in \mathcal{N}$, let state matrix $x$ define as $x(t_k) = \begin{bmatrix} X(t_k) & \overrightarrow{V}(t_k) \end{bmatrix}^T$ with real location $X$ and velocity $\overrightarrow{V}$, then $x(t_k)$ denotes the actual state at time $t_k$. In the same way, we define the measurement state $x^{'}(t_k)$ at time $t_k$ as $x^{'}(t_k) = \begin{bmatrix} X^{'}(t_k) & \overrightarrow{V^{'}}(t_k)\end{bmatrix}^{T}$.

During time interval $\Delta T$, which is the elapsed time from the previous updated time $t_{k-1}$ until current time $t_k$, i.e., $\Delta T = t_k - t_{k-1}$, the node moves from $X(t_{k-1})$ to $X(t_{k})$ such that $X(t_k) = X(t_{k-1}) + \Delta T \overrightarrow{V}(t_k)$. Hence, the measured velocity $\overrightarrow{V}^{'}_{i}(t_k)$ is
\begin{eqnarray}\label{eq:measuredspeed}
\overrightarrow{V}^{'}_{i}(t_k) &=& \frac{X_{i}^{'}(t_k) - X_{i}^{'}(t_{k-1})}{\Delta T}\nonumber \\
&=& \overrightarrow{V}_{i}(t_k) + \frac{1}{\Delta T}  W_{i}(t_k, t_{k-1}) \nonumber
\end{eqnarray}
where $W_{i}(t_k, t_{k-1})$ is the sum of measurement errors at times $t_k$ and $t_{k-1}$.

Suppose that during elapsed time $\Delta T$ the velocity is constant, i.e., $\overrightarrow{V}(t_k) = \overrightarrow{V}(t_{k-1})$. The actual state $x(t_k)$ and measurement state $x^{'}(t_k)$ can be written as
\begin{eqnarray} \label{eq:KF0}
    x(t_k)     &=& \begin{bmatrix} 1 & \Delta T \\ 0 & 1\end{bmatrix} x(t_{k-1}) \nonumber \\
    x^{'}(t_k) &=& \begin{bmatrix} 1 & 0 \\ 0 & 1\end{bmatrix} x(t_{k}) + w(t_k), \nonumber
\end{eqnarray}
where  $w(t_k) = \begin{bmatrix} W(t_k) & \frac{1}{\Delta T}  W(t_k, t_{k-1}) \end{bmatrix}^T$. Let denote matrix $A(t_{k-1}) = \begin{bmatrix} 1 & \Delta T \\ 0 & 1 \end{bmatrix}$ and matrix $B = \begin{bmatrix} 1 & 0  \\ 0 & 1 \end{bmatrix}$. The matrix $A(t_{k-1})$ represents the state change and the matrix $B$ describes the relation between the actual state and measurement state. The above equation can be rewritten as
\begin{eqnarray} \label{eq:KF1}
    x(t_k) &=& A(t_{k-1}) x(t_{k-1}) \nonumber \\
    x^{'}(t_k) &=& B x(t_k) + w(t_k). \nonumber
\end{eqnarray}

Since the actual state $x(t_k)$ cannot directly be acquired, we define $\hat{x}^{-}(t_k)$ as $a~priori$ estimate at time $t_k$ for a given state prior to time $t_k$, and ${\hat{x}}(t_k)$ as $a~posteriori$ estimate state at time $t_k$ for a given measurement state $x^{'}(t_k)$. Let $P^{-}(t_k)$ and $P(t_k)$ be $a~priori$ estimate error covariance and $a~posteriori$ estimate error covariance, respectively, and can be expressed by
 \begin{eqnarray} \label{eq:KF2}
    P^{-}(t_k) &=& E \Big[\big( x(t_k)-\hat{x}^{-}(t_k)\big) \big(x(t_k)-\hat{x}^{-}(t_k)\big)^T \Big] \nonumber \\
    P(t_k) &=& E \Big[\big( x(t_k)-\hat{x}(t_k) \big) \big( x(t_k)-\hat{x}(t_k) \big)^T \Big]. \label{aposteriori}
\end{eqnarray}

To find the best estimate of the current state, we apply the Kalman filter. The operation of the Kalman filter includes two mechanisms: time update and measurement update. The time update process is responsible to predicting the current estimate state based on the previous state by computing ${\hat{x}}^{-}(t_k)$ and $P^{-}(t_k)$
\begin{eqnarray}\label{eq:TimeUpdate}
   {\hat{x}}^{-}(t_k) &=& A_{t_{k-1}}~\hat{x}(t_{k-1}), \nonumber \\
   P^{-}(t_k) &=& A(t_{k-1}) ~ P(t_{k-1}) ~ A^{T}(t_{k-1}).\nonumber
\end{eqnarray}

\begin{figure}
  \centering
  \includegraphics[scale=0.5]{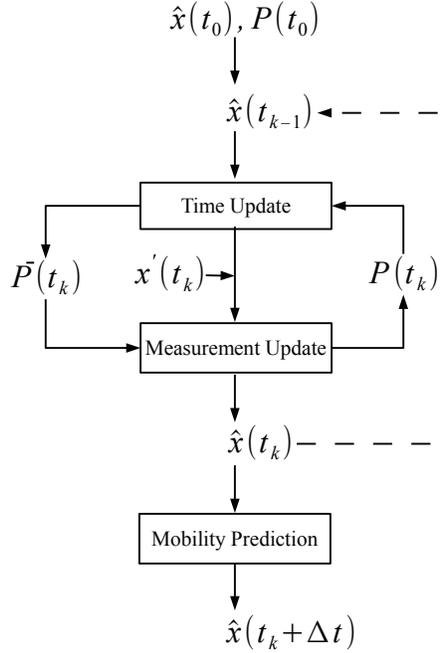}
  \caption{The Kalman filter based location correction process}\label{lb:proposedAlgorithm}
\end{figure}

After the time update operation, the measurement update corrects the measurement state as follow:
 \begin{eqnarray} \label{eq:MeasurementUpdate}
    K(t) &=& P^{-}(t_k) B^{T} \big( B P^{-}(t_k) B^{T} + R \big)^{-1} \nonumber \\
    {\hat{x}}(t_k) &=& {\hat{x}}^{-}(t_k) + K(t_k) \big(x^{'}(t_k) - B {\hat{x}}^{-}(t_k)\big) \nonumber \\
    P(t_k) &=& \big(I - K(t_k) B \big) P^{-}(t_k), \nonumber
 \end{eqnarray}
where $K(t_k)$ and $R$ are the Kalman gain and the measurement error covariance, respectively.
After that, the operation is repeated, the estimate state is measured based on the previous state and measurement state. Each node updates and tracks its current location based on periodically or eventually measured locations as the process of the discrete Kalman filter algorithm, which is summarized in Fig.~\ref{lb:proposedAlgorithm}.

In implementation, the measurement error covariance $R$ is measured prior to the operation of the Kalman filter. The measurement error covariance is determined by the variance of measurement noise by obtaining some off-line sample measurement~\cite{greg2006}. The initial value for each state $\hat{x}(t_0)$ is set to the measured information at the beginning.

In addition, we can obtain the confidence level of a link duration from the $a~posteriori$ estimate error covariance matrix~$P(t_k)$. The $a~posteriori$ estimate error covariance matrix in~(\ref{aposteriori}) can be reexpressed as
 \begin{eqnarray} \label{eq:EstimationCov}
    P(t_k)
    &=&  \begin{bmatrix} E\left[ e_{X}^{2}(t_{k})\right] & E\left[\frac{1}{\Delta T} \left( e_{X}^{2}(t_{k}) - e_{X}(t_{k})e_{X}(t_{k-1}) \right)\right] \\ E\left[\frac{1}{\Delta T} \left( e_{X}^{2}(t_{k}) - e_{X}(t_{k})e_{X}(t_{k-1}) \right)\right] & E\left[ \frac{1}{\Delta T^{2}} \left(\left( e_{X}(t_{k}) - e_{X}(t_{k-1}) \right) \right)^{2}\right] \end{bmatrix}, \nonumber
 \end{eqnarray}
where $e_{X}(t_{k}) \equiv X(t_k)-\hat{X}(t_k)$. The square root of the expected square error~$E\left[ e_{X}^{2}(t_{k})\right]$ is equivalently considered as the standard deviation in the engineering community~\cite{rms}. Hence, the root-mean-square error, $\sqrt{E\left[ e_{X}^{2}(t_{k})\right]}$, is equivalently the standard deviation of errors, and $\frac{\sqrt{E\left[ e_{X_{i}}^{2}(t_{k})+e_{X_{j}}^{2}(t_{k})\right]}}{\overrightarrow{V}_{(i,j)}}$, denoted as $\varepsilon$, becomes the confidence level of link duration of link $(i, j)$.

\subsection{The Enhanced Mobility Prediction Routing Protocol}
\label{MobilityPrediction}
In this subsection, we develop a mobility prediction-based routing protocol in the presence of location errors. Our goal of mobility prediction is to find the longest RET and to avoid the risky link. The risky link is defined as a link with vulnerable link duration time LDT seems to be dead or to be no longer alive in a short time after discovering.

When new data arrive at a node, the source node finds an active route associated with the corresponding destination in its routing table, as in Subsection~\ref{Routing-Protocol}. If no active route exists, the source node initiates route discovery to find a route to the destination node by broadcasting a RREQ message with recently updated location information and the standard deviation $\sqrt{E\left[ e_{X}^{2}(t_{k})\right]}$ of location error to neighboring nodes. The RET field and the hop count field in the RREQ message are initially set to infinity and one, respectively.

Upon reception of RREQ, a node computes the link duration time between the RREQ sender and itself, which implies the estimated lifetime of the link, from~(\ref{eq-ldt}). To compute link durations in~\ref{eq-ldt}, nodes use the compensated location information $\hat{x}(t_k)$ instead of the measured location information $x^{'}(t_k)$. To exclude the risky link, the node compares LDT value with the confidence level $\varepsilon$ of LDT, which is computed from the standard deviations of the RREQ sender and itself. If the LDT value is less than $\varepsilon$, the node discards the RREQ. Otherwise, the LDT value updates a RET value in the RREQ. If the LDT is smaller than the RET in the RREQ, the receiving node replace the RET value by the new LDT. If the RREQ receiver is not the destination of the RREQ, the node broadcasts the receiving RREQ to other nodes after increasing the hop count by one until the RREQ reaches the destination.

\begin{figure}
  \centering
  \includegraphics[width=2.65in]{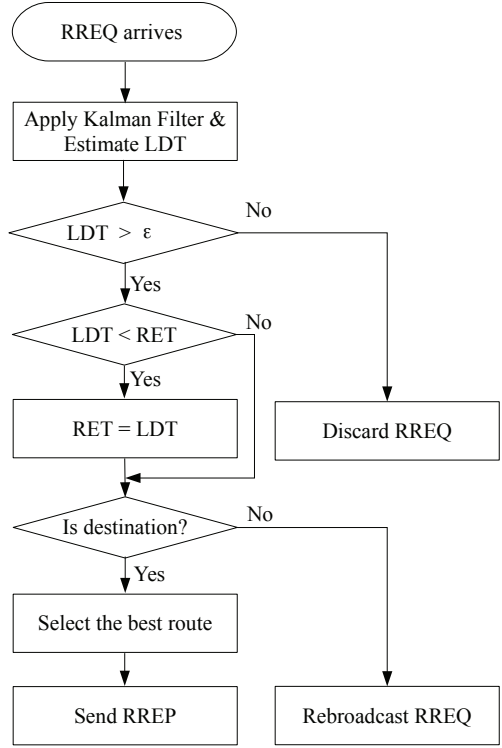}
  \caption{The Kalman filter based enhanced mobility prediction: EMP}\label{NodeActions}
\end{figure}
In the case when a node is the destination of RREQ, the node waits for time interval~$T_{w}$ and collects RREQs whose destination is the node. After the time interval~$T_{w}$, the destination chooses the longest route among the received routes and replies a RREP message after setting the lifetime field as the corresponding RET. RREP receivers relay the RREP message in a unicast manner until the RREP reaches the source, as described in Subsection~\ref{Routing-Protocol}. The details of proposed algorithm, AODV with enhanced mobility prediction~(EMP), are described in Fig.~\ref{NodeActions}.

For route maintenance, we adopt the adaptive period for hello messages as in~\cite{Chao07, Cons2010}, referred to as hello interval adjustment~(HIA), to reduce the overheads instead of a fixed period. When receiving a RREQ from node~$i$, node~$j$ estimates link duration $LDT_{(i,j)}$ in Fig.~\ref{NodeActions}, and set the period for hello frequency to
\[\max\left\{T_{\min},\frac{\min_{i \in {\cal N}_{j}}LDT_{(i,j)}}{\beta} \right\},\]
where $T_{\min}$ is the minimum value for the hello period, ${\cal N}_{j}$ is a set of the nodes that establish active links with node~$j$, and $\beta$ is a control parameter. The value of $\beta$ is greater than or equal to 1, which aims to adjust the hello frequency.

\section{Performance Evaluation}
\label{Evaluation}
We evaluate the performance of our proposed algorithms by using the network simulator NS-2 \cite{V03}. For simulations, there are 100 nodes initially distributed in an area of 2~km by 1.5~km and the transmission range of each node is set to 250~m. We run simulations with ten different random seeds, and average the simulation results.

\begin{table}
\caption{Parameter settings} 
\centering
\begin{tabular}{l l} 
\hline 
Parameter & Values \\ 
\hline 
Network simulator                   & NS-2.34 \\ 
Simulation area                     & 2 km $\times$ 1.5 km \\ %
Number of mobile nodes              & 100 \\
Simulation time                     & 900 s \\
Mobility model                      & Random way point \\
Pause time                          & 0 s\\
Packet generation rate              & 4 packets/s \\
Packet size                         & 512 bytes \\
Transmission range                  & 250 m \\ 
\hline 
\end{tabular}
\label{parameter} 
\end{table}
The Random Waypoint Mobility (RWP)~\cite{hy05} model is used as a referenced mobility model, in which mobile nodes move from their current locations to new locations by randomly choosing directions and speeds. Upon arrival at a destination, after a pause time, they choose another random destinations in the simulation area and travel toward the destinations with a uniformly distributed speed between the maximum speed and minimum speed. We set the pause time to zero to represent constant mobility.

The constant bit rate (CBR) traffic under the user datagram protocol (UDP) is used to accurately compare different routing protocols with a sending rate of 4 packets per second and 512 bytes of packet size. The parameter settings are listed in Table ~\ref{parameter}.

Two metrics are used for evaluating the network performance: the packet delivery rate and the normalized routing load. The packet delivery ratio is defined as the ratio of the number of generated packets to the number of packets received at the corresponding destinations. For the amount of overhead packets, we count the number of packets used for route discovery and route maintenance. For comparison, the total number of overhead packets is normalized by the number of packets successfully delivered to destinations.

To evaluate the performance improvement, our EMP routing protocol is compared with mobility prediction-based AODV routing protocol with route discovery mechanism~\cite{HU11} and the conventional AODV routing protocol in various noisy environments. For simplicity, the mobility prediction-based AODV routing protocol is denoted as the classic mobility prediction (MP). For simulations, we assume that the location errors of each node $i$ are Gaussian random variables with zero mean and standard deviation $\sigma _i$.

Firstly, we compare the performance of our enhanced mobility prediction EMP with the previous work MP in the presence of location errors by varying the standard deviation of location errors from $3~m$~($1.12~\%~$of transmission range) to $50~m$~($20~\%~$of transmission range).

Secondly, we fix the standard deviation of location errors to $20~m~$($8~\%~$of transmission range) and show the network performance under different impact of network environments, such as the impact of node velocity, traffic load, and node density. For each scenario, the HIA mechanism is enabled or disabled to show the impact of adaptive hello period.

\subsection{The Performance of the Kalman Filter based Enhanced Mobility Prediction in the Presence of Location Errors}
\label{EMP}
To compare our EMP routing protocol with the MP routing protocol, ten source-destination pairs generate 4 packets per second during the simulation time. For mobility, each node follows the RWP mobility model with randomly selected speed between 1~m/s and 20~m/s.

Our proposal incorporates the Kalman filter to remove the location errors in order to reduce the impact of location errors, and predicts the link duration time more accurately. The EMP can also improve the network performance by limiting the number of route discovery due to the dangerous link with an uncertain link duration time. The node establishing the uncertain link duration time does not allow to forward the RREQ messages. Therefore, the discovered route becomes a better candidate for route selection and the number of overhead messages is significantly decreased.

\begin{figure}
        \centering
        \begin{subfigure}[b]{0.522\textwidth}
             \centering
            \includegraphics[width=\textwidth]{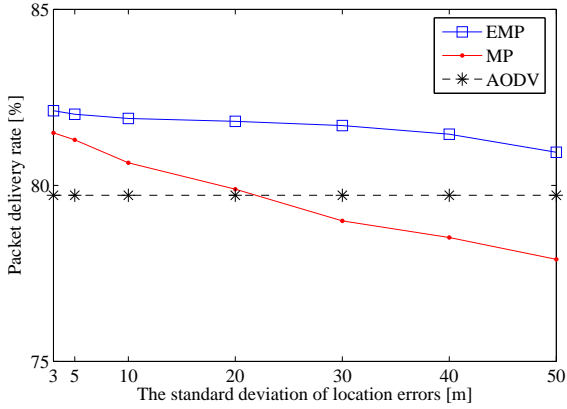}
            \caption{Packet delivery rate versus location errors}\label{LE-PDR}
        \end{subfigure}%
        \begin{subfigure}[b]{0.522\textwidth}
            \centering
            \includegraphics[width=\textwidth]{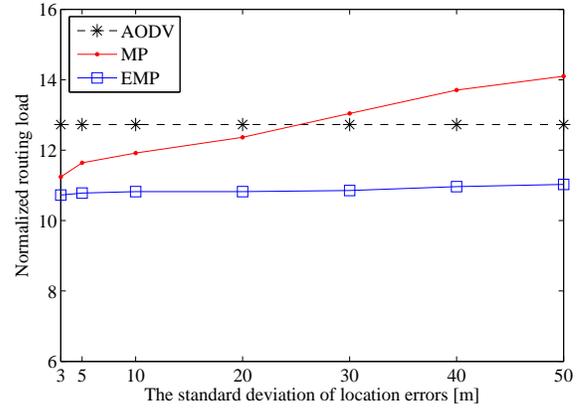}
            \caption{Normalized routing load versus location errors }\label{LE-NRL}
        \end{subfigure}
        \caption{Impact of location errors - fixed hello interval}\label{LocationErrors_noHIA}
\end{figure}

In Fig.~\ref{LE-PDR}, the packet delivery rates of EMP, MP, and AODV routing protocols are compared. As the standard deviation of location errors increases, the packet delivery rate of the MP routing protocol is decreased faster than that of EMP. When the standard deviation of location errors is behind a certain level~( 20~m in this case), the packet delivery rate of the MP routing protocol is lower than that of the AODV routing protocol. The large location errors lead to poor mobility prediction, which results in performance degradation. However, the packet delivery rate of our proposed routing protocol EMP outperforms MP and AODV routing protocols in all the cases and is robust to the location errors.

Fig.~\ref{LE-NRL} shows the normalized routing loads of EMP, MP, and AODV routing protocols. As the standard deviation of location errors increases, the normalized routing loads of MP and EMP increases due to inaccurate prediction. The normalized routing load of MP increases faster than that of EMP and is even greater than that of the conventional AODV. However, the EMP just slightly increases the routing overhead, which demonstrates that our proposed algorithm is robust to location errors.

\begin{figure}
        \centering
        \begin{subfigure}[b]{0.522\textwidth}
        \centering
        \includegraphics[width=\textwidth]{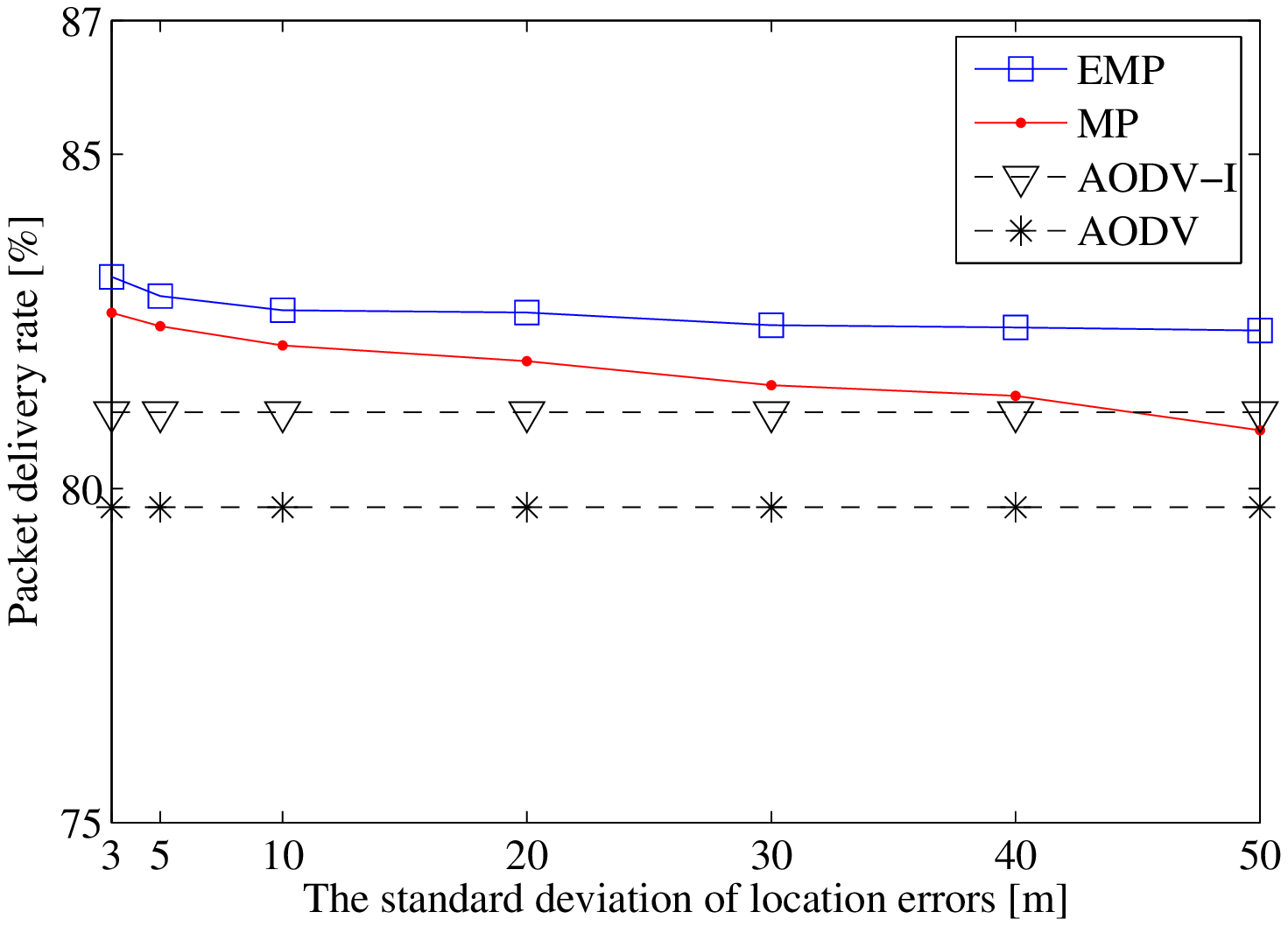}
        \caption{Packet delivery rate versus location errors}\label{LE-HIA-PDR}
        \end{subfigure}%
        \begin{subfigure}[b]{0.522\textwidth}
  \centering
        \includegraphics[width=\textwidth]{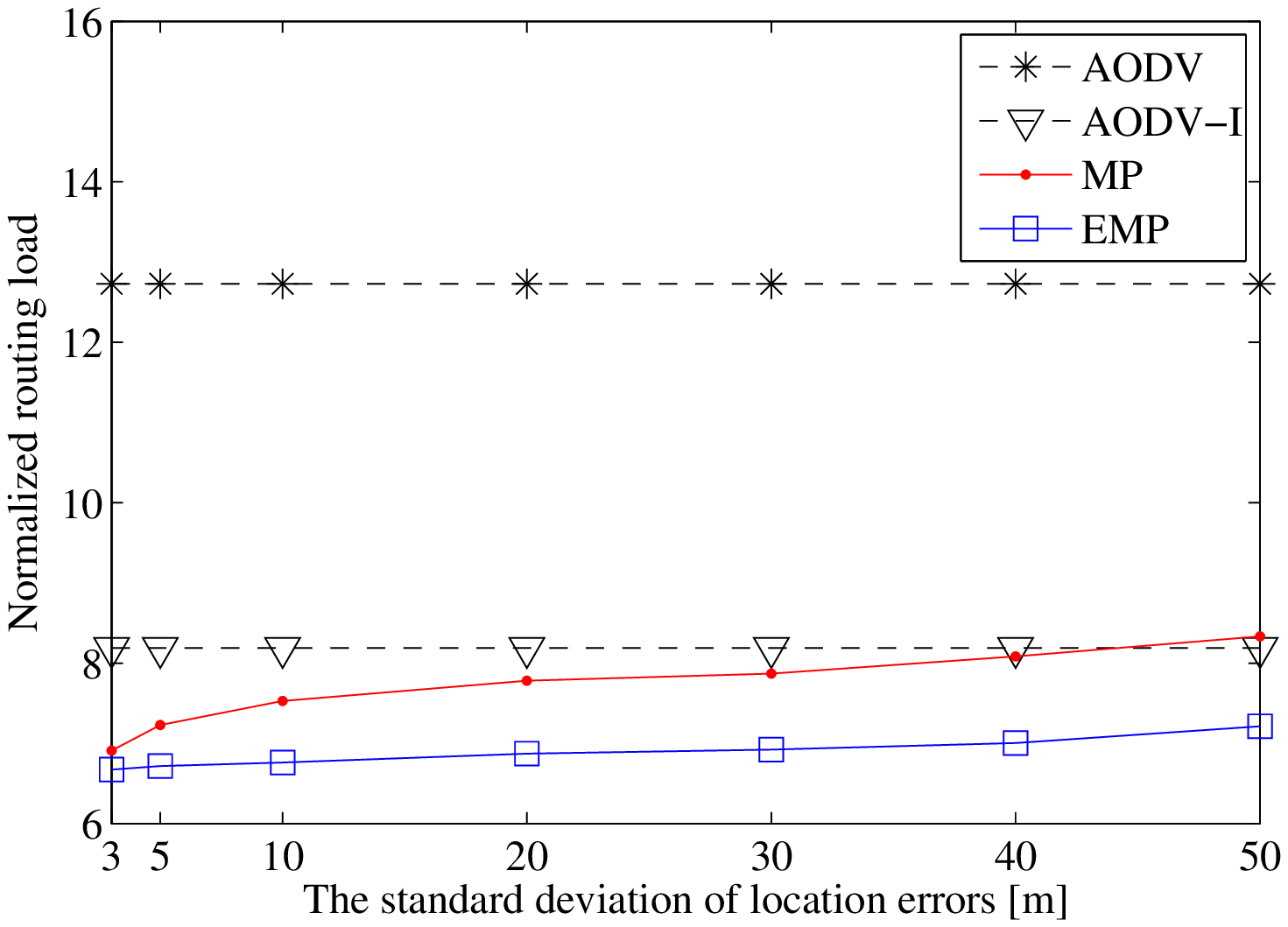}
        \caption{Normalized routing load versus location errors}\label{LE-HIA-NRL}
        \end{subfigure}
        \caption{Impact of location errors - flexible hello interval}\label{LocationErrors_HIA}
        \end{figure}

Figs.~\ref{LE-HIA-PDR} and~\ref{LE-HIA-NRL} show the packet delivery rate and the normalized routing load when the HIA is enabled for the mobility prediction-based routing protocol. The HIA mechanism is used for reducing the unnecessary hello messages. The AODV routing protocol sets the hello frequency to 1 second and the AODV-I sets the hello frequency to 20 seconds. As the location errors increases, the performance of MP is degraded. It is because the MP routing cannot estimate the true value of link duration that leads to incorrect route selection. Therefore, the selected route is unreliable and unstable so that the source node has to handle the route more frequently. When the standard deviation of location errors is larger than $40~m$, the performance of the MP routing is lower than the AODV-I routing. The inaccurate link duration for selecting the route and setting the hello interval causes the performance degradation of mobility prediction-based routing protocol without location error compensation.

\subsection{The Impact of Node Velocity}
\label{IP:NodeVelocity}
We study the impact of node velocity on routing performance in various network environments. The node mobility has a great impact on network performance \cite{Wu09, Amjad10} since the change of topology leads to more exchanging messages in order to find and maintain new routes. During simulations, performances are compared in three different mobile environments: low mobility, medium mobility, and high mobility. For the low mobility environment, we set the speed for RWP to 1~m/s, which is a pedestrian speed~(3.6~km/h). We also set 10~m/s and 20~m/s~(72~km/h) as the node speeds for the medium mobility and the high mobility environments, respectively.

\begin{figure}
        \centering
        \begin{subfigure}[b]{0.522\textwidth}
                \centering
                \includegraphics[width=\textwidth]{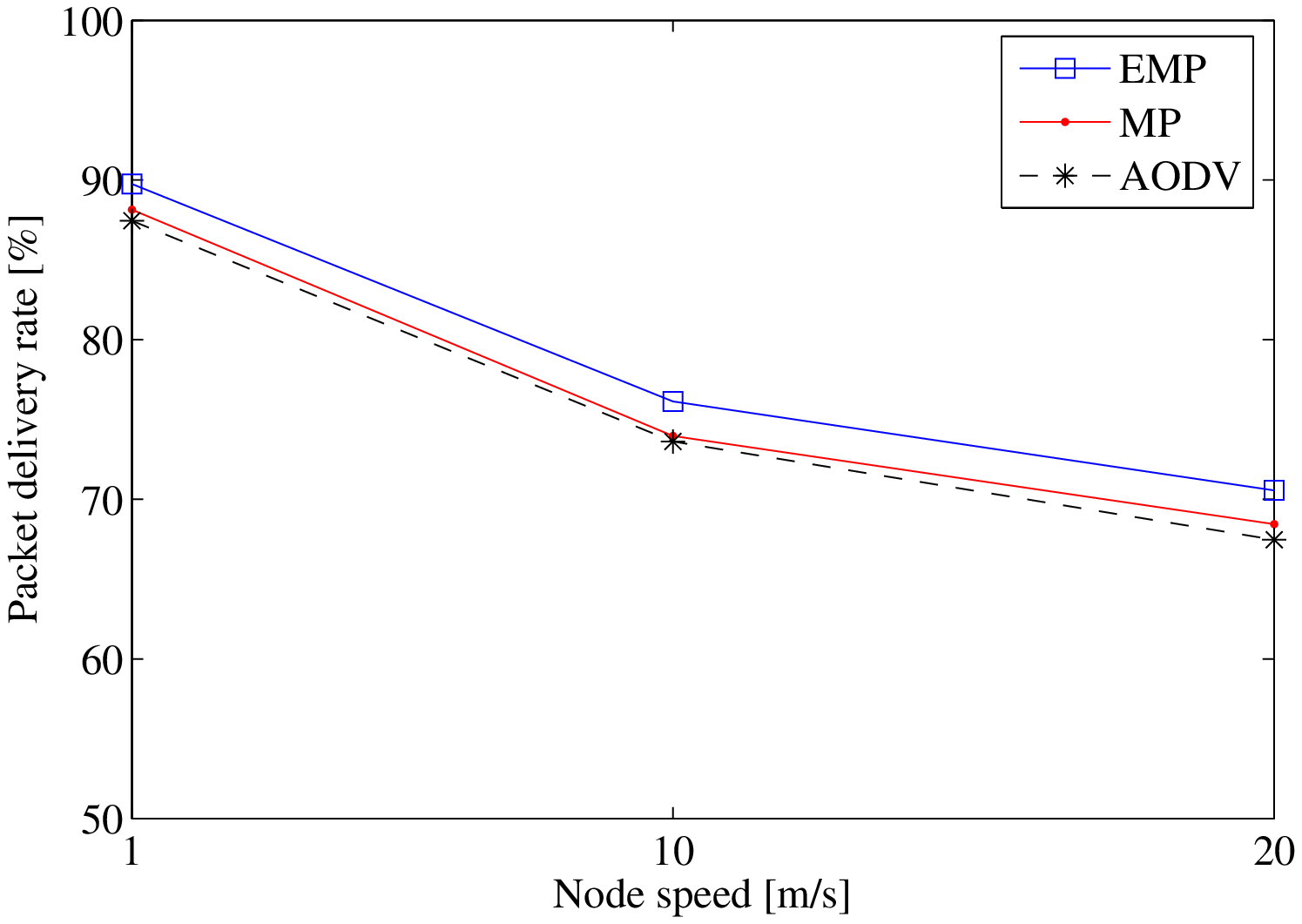}
                \caption{fixed hello interval}
                \label{NodeVelocity_PDR_noHIA}
        \end{subfigure}%
        \begin{subfigure}[b]{0.522\textwidth}
                \centering
                \includegraphics[width=\textwidth]{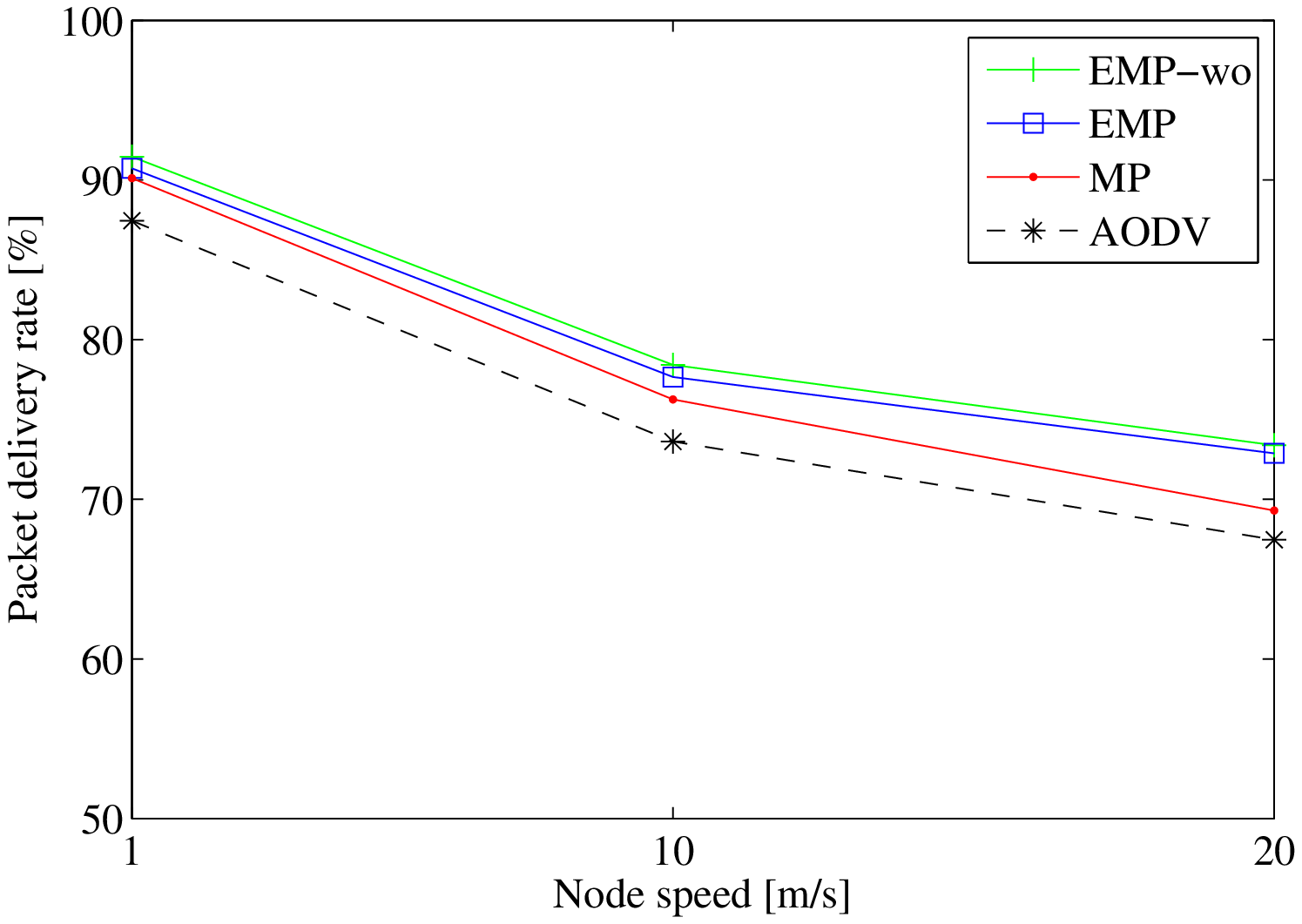}
                \caption{flexible hello interval}
                \label{NodeVelocity_PDR_HIA}
        \end{subfigure}
        \caption{Packet delivery rate versus node velocity}\label{NodeVelocity_PDR}
\end{figure}

\begin{figure}
        \centering
        \begin{subfigure}[b]{0.522\textwidth}
                \centering
                \includegraphics[width=\textwidth]{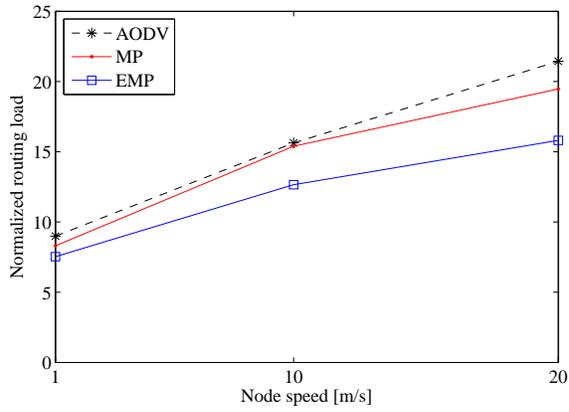}
                \caption{fixed hello interval}
                \label{fig:NodeVelocity_NRL_noHIA}
        \end{subfigure}%
        \begin{subfigure}[b]{0.522\textwidth}
                \centering
                \includegraphics[width=\textwidth]{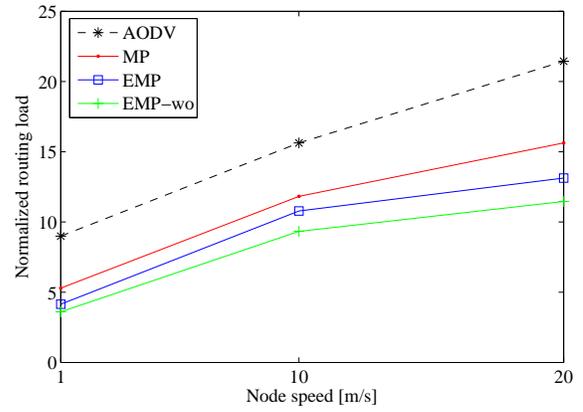}
                \caption{flexible hello interval}
                \label{fig:NodeVelocity_NRL_HIA}
        \end{subfigure}
        \caption{Normalized routing load versus node velocity}\label{fig:NodeVelocity_NRL}
\end{figure}

Fig.~\ref{NodeVelocity_PDR} shows that the packet delivery rate decreases as the node velocity increases since routes are more frequently broke and more overhead messages are necessary due to fast topology change, as shown in Fig.~\ref{fig:NodeVelocity_NRL}. Whether hello interval for route maintenance is fixed or adaptive to mobility, AODV with mobility prediction is better than the conventional AODV in the presence of location errors, as shown in~\cite{su2001}. Our algorithm, which compensates for location errors, outperforms the others and is close to the case (EMP-wo) when location information is error-free. Therefore, our proposed routing protocol EMP can adapt to the scalability network even in the presence of location errors.

\subsection{The Impact of Traffic Load}
\label{IP:TrafficLoad}
Traffic load can affect the performance of routing protocols. To study the impact of traffic load, we vary the number of source-destination pairs to deliver generated data. For mobility, each node also follows the RWP mobility model with randomly selected speed between 1~m/s and 20~m/s.

\begin{figure}
        \centering
        \begin{subfigure}[b]{0.522\textwidth}
                \includegraphics[width=\textwidth]{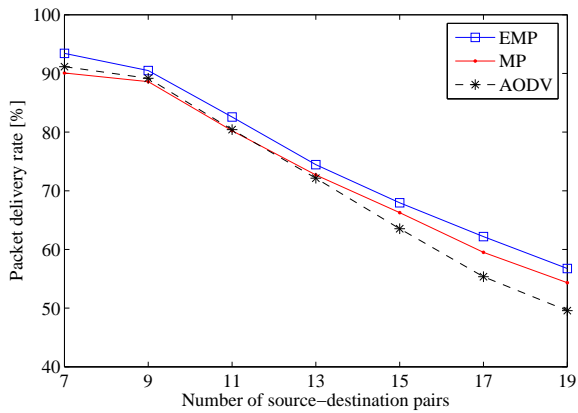}
                \caption{fixed hello interval}
                \label{fig:TrafficLoad_PDR_noHIA}
        \end{subfigure}%
        \begin{subfigure}[b]{0.522\textwidth}
                \includegraphics[width=\textwidth]{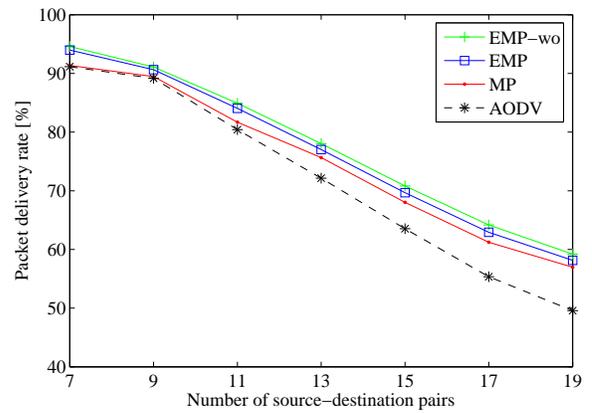}
                \caption{flexible hello interval}
                \label{fig:TrafficLoad_PDR_HIA}
        \end{subfigure}
        \caption{Packet delivery rate versus traffic load}\label{fig:TrafficLoad_PDR}
\end{figure}

Fig.~\ref{fig:TrafficLoad_PDR} shows the packet delivery rates. As increase of the number of source-destination pairs in the network, due to transmission collision and congestion, the packet delivery rates are reduced. In Figs.~\ref{fig:TrafficLoad_PDR_noHIA} and~\ref{fig:TrafficLoad_PDR_HIA}, our algorithm outperforms the others and almost closed to the EMP-wo, which assumes no location errors in measurement and is an upper bound of the performance. That means that our proposed algorithm EMP is robust to the location errors.

\begin{figure}
        \centering
        \begin{subfigure}[b]{0.522\textwidth}
                \includegraphics[width=\textwidth]{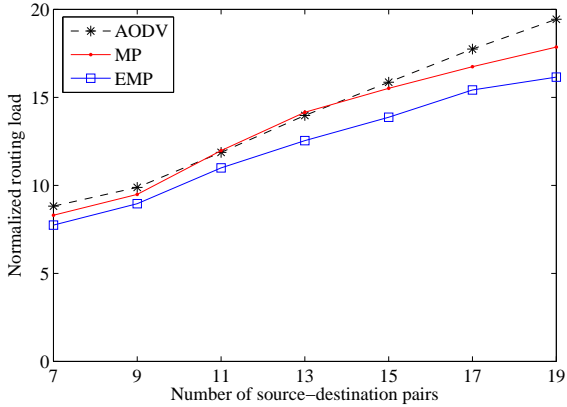}
                \caption{fixed hello interval}
                \label{fig:TrafficLoad_NRL_noHIA}
        \end{subfigure}%
        \begin{subfigure}[b]{0.522\textwidth}
                \includegraphics[width=\textwidth]{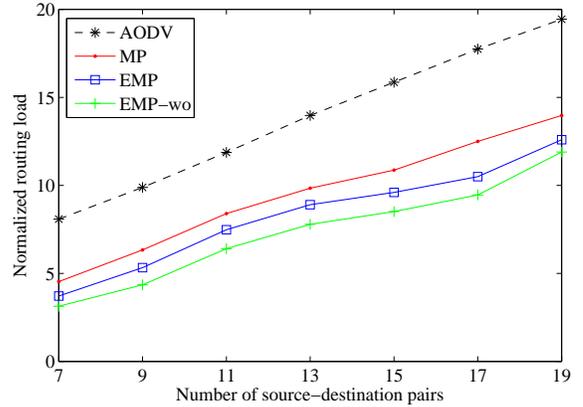}
                \caption{flexible hello interval}
                \label{fig:TrafficLoad_NRL_HIA}
        \end{subfigure}
        \caption{Normalized routing load versus traffic load}\label{fig:TrafficLoad_NRL}
\end{figure}
Fig.~\ref{fig:TrafficLoad_NRL} reports the normalized routing load when increasing the traffic load. In Fig.~\ref{fig:TrafficLoad_NRL_noHIA}, the HIA mechanism is disabled, the MP needs to exchange more routing messages caused by the location errors, whereas the EMP can reduce the amount of routing overhead as compared to the MP and the original AODV. When the HIA mechanism is enabled, a large number of hello messages are reduced, but the hello message still contributes well to the local connectivity management. The EMP and EMP-wo routing protocol can sharply reduce a great number of overhead as compared with the MP and the original AODV routing protocol.

\subsection{The Impact of Node Density}
\label{IP:NodeDensity}
\begin{figure}
        \centering
        \begin{subfigure}[b]{0.522\textwidth}
                \includegraphics[width=\textwidth]{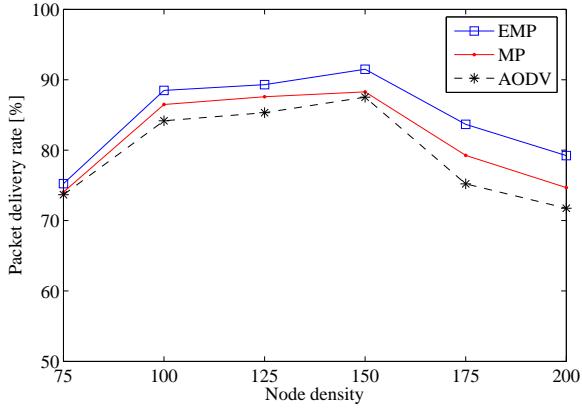}
                \caption{fixed hello interval}
                \label{fig:NodeDensity_PDR_noHIA}
        \end{subfigure}%
        \begin{subfigure}[b]{0.522\textwidth}
                \includegraphics[width=\textwidth]{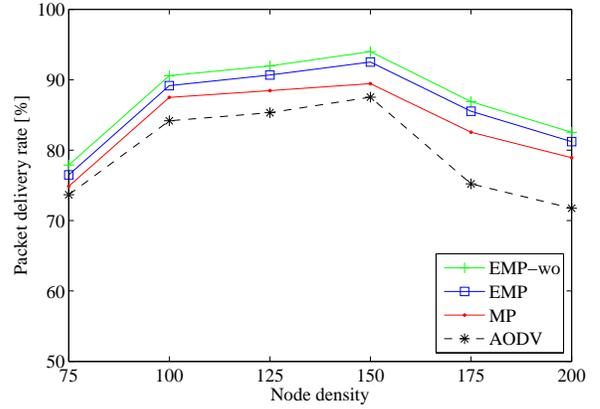}
                \caption{flexible hello interval}
                \label{fig:NodeDensity_PDR_HIA}
        \end{subfigure}
        \caption{Packet delivery rate versus node density}\label{fig:NodeDensity_PDR}
\end{figure}

\begin{figure}
        \centering
        \begin{subfigure}[b]{0.522\textwidth}
                \includegraphics[width=\textwidth]{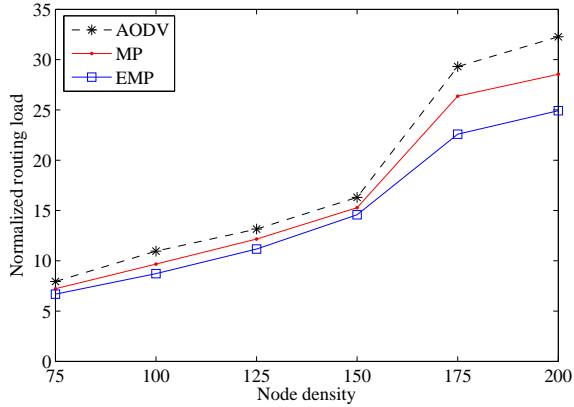}
                \caption{fixed hello interval}
                \label{fig:NodeDensity_NRL_noHIA}
        \end{subfigure}%
        \begin{subfigure}[b]{0.522\textwidth}
                \includegraphics[width=\textwidth]{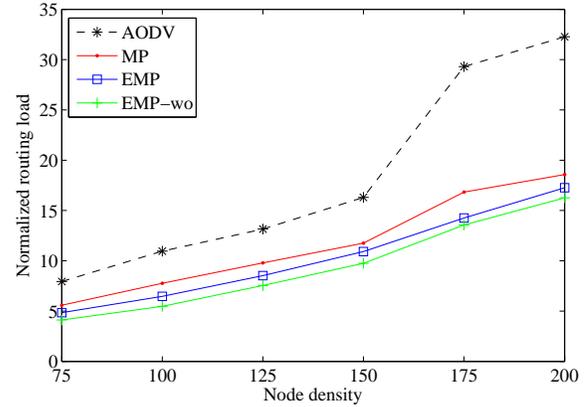}
                \caption{flexible hello interval}
                \label{fig:NodeDensity_NRL_HIA}
        \end{subfigure}
        \caption{Normalized routing load versus node density}\label{fig:NodeDensity_NRL}
\end{figure}

In this subsection, we study the impact of node density on routing performance by varying the number of nodes from 75 nodes to 200 nodes as shown in Fig.~\ref{fig:NodeDensity_PDR} and Fig.~\ref{fig:NodeDensity_NRL}. If the number of nodes is too small, feasible routes between sources and destinations may not exist in the network so that the routing performance improves as the number of nodes increase in the network. However, above a certain number of nodes, the larger number of node hinders packet delivery due to larger overhead messages required to maintain and discover routes. The EMP still outperforms the MP with respect to the packet delivery rate and the overhead in the presence of location errors.

\section{Conclusion}
\label{Conclusion}
This paper proposed an on-demand routing algorithm with enhanced mobility prediction that takes into account the location errors. Imperfect location information induces the performance degradation, but location errors in measurement were ignored in previous work. In the presence of location errors, we develop an on-demand routing algorithm collaborating to the Kalman filter to predict node mobility. Since the Kalman filter provides the root-mean-square error between the actual location and estimated location, the proposed algorithm exclude unreliable links considering the confidence levels of links. The estimated link duration adapt to the route maintenance period to reduce overheads. Via simulations, our proposed algorithm is robust to location errors and outperforms the previous algorithms.

\section*{Acknowledgment}
The authors would like to acknowledge Dang Quoc Khanh for helpful discussions about the Kalman Filter theory on the paper.




\bibliographystyle{elsarticle-num}
\bibliography{mybibdatabase}

\begin{thebibliography}{10}
\expandafter\ifx\csname url\endcsname\relax
  \def\url#1{\texttt{#1}}\fi
\expandafter\ifx\csname urlprefix\endcsname\relax\def\urlprefix{URL }\fi
\expandafter\ifx\csname href\endcsname\relax
  \def\href#1#2{#2} \def\path#1{#1}\fi

\bibitem{Ismail07a}
D.~Ismail, M.~Jaafar, Mobile ad hoc network overview, in: Asia-Pacific
  Conference on Applied Electromagnetics, 2007, pp. 1--8.

\bibitem{Corson99a}
S.~Corson, J.~Macker, Rfc 2501 routing protocol performance issues and
  evaluation considerations on mobile ad hoc networking (1999).

\bibitem{Camp02a}
T.~Camp, J.~Boleng, V.~Davies, A survey of mobility models for ad hoc network
  research, Wireless Communications and Mobile Computing: special issue on
  Mobile Ad hoc Networking: research, trends and applications 2 (2002) 483 --
  502.

\bibitem{Bai03a}
F.~Bai, N.~Sadagopan, A.~Helmy, A framework to systematically analyze the
  impact of mobility on performance of routing protocols for adhoc networks,
  in: INFOCOM, Vol.~2, 2003, pp. 825 -- 835.

\bibitem{P99ad}
C.~Perkins, E.~Royer, Ad-hoc on-demand distance vector routing, in: Second IEEE
  Workshop on Mobile Computing Systems and Applications, Proceedings, 1999, pp.
  90 --100.

\bibitem{Tsao06}
C.-L. Tsao, Y.~eh~T~Wu, W.~Liao, J.-C. Kuo, Link duration of the random way
  point model in mobile ad hoc networks, in: IEEE Wireless Communications and
  Networking Conference, Vol.~1, 2006, pp. 367 --371.

\bibitem{Jiang05}
S.~Jiang, D.~He, J.~Rao, A prediction-based link availability estimation for
  routing metrics in manets, IEEE/ACM Transactions on Networking 13~(6) (2005)
  1302 -- 1312.

\bibitem{Luo10}
H.~Luo, D.~Laurenson, Link-duration-oriented route lifetime computation for
  aodv in manet, in: International Conference on Wireless Communications and
  Signal Processing, 2010, pp. 1 --4.

\bibitem{Fan04h}
F.~Bai, N.~Sadagopan, B.~Krishnamachari, A.~Helmy, Modeling path duration
  distributions in manets and their impact on reactive routing protocols, IEEE
  Journal on Selected Areas in Communications 22~(7) (2004) 1357--1373.

\bibitem{Namuduri12}
K.~Namuduri, R.~Pendse, Analytical estimation of path duration in mobile ad hoc
  networks, IEEE Sensors Journal 12~(6) (2012) 1828 --1835.

\bibitem{La07}
R.~La, Y.~Han, Distribution of path durations in mobile ad hoc networks and
  path selection, IEEE/ACM Transactions on Networking 15~(5) (2007) 993 --1006.

\bibitem{su2001}
W.~Su, S-J.Lee, M.~Gerla, Mobility prediction and routing in ad hoc wireless
  networks, International Journal of Network Management 11~(1) (2001) 3 --30.

\bibitem{HU11}
X.~Hu, J.~Wang, C.~Wang, Mobility-adaptive routing for stable transmission in
  mobile ad hoc networks, Journal of Communications 6~(1) (2011) 79--86.

\bibitem{Chao07}
L.~Chao, H.~Aiqun, Reducing the message overhead of aodv by using link
  availability prediction, in: Proceedings of the 3rd International Conference
  on Mobile Ad-hoc and Sensor Networks, MSN'07, Springer-Verlag, 2007, pp.
  113--122.

\bibitem{skwon06a}
S.~Kwon, N.~Shroff, Geographic routing in the presence of location errors,
  Computer Networks 50 (2006) 2902 --2917.

\bibitem{misra2011}
P.~Misra, P.~Enge, Global positioning system: Signals, measurements and.

\bibitem{Drawil13}
N.~Drawil, H.~Amar, O.~Basir, Gps localization accuracy classification: A
  context-based approach, IEEE Transactions on Intelligent Transportation
  Systems 14~(1) (2013) 262--273.

\bibitem{greg2006}
G.~Welch, G.~Bishop, An introduction to the kalman filter, Department of
  Computer Science, University of North Carolina at Chapel Hill, NC.

\bibitem{rms}
E.~W. Weisstein, Standard deviation, from MathWorld--A Wolfram Web Resource.

\bibitem{Cons2010}
N.~Hernandez-Cons, S.~Kasahara, Y.~Takahashi, Dynamic hello/timeout timer
  adjustment in routing protocols for reducing overhead in \{MANETs\}, Computer
  Communications 33~(15) (2010) 1864 -- 1878.

\bibitem{V03}
K.~Fall, K.~Varadhan, The ns manual (formerly ns notes and documentation)
  (2005).

\bibitem{hy05}
E.~Hyyti{\"a}, H.~Koskinen, P.~Lassila, A.~Penttinen, J.~Roszik, J.~Virtamo,
  Random waypoint model in wireless networks, Networks and Algorithms:
  complexity in Physics and Computer Science, Helsinki.

\bibitem{Wu09}
Y.-T. Wu, W.~Liao, C.-L. Tsao, T.-N. Lin, Impact of node mobility on link
  duration in multihop mobile networks, IEEE Transactions on Vehicular
  Technology 58~(5) (2009) 2435 --2442.

\bibitem{Amjad10}
K.~Amjad, A.~Stocker, Impact of node density and mobility on the performance of
  aodv and dsr in manets, in: 7th International Symposium on Communication
  Systems Networks and Digital Signal Processing, 2010, pp. 61 --65.

\end{thebibliography}
\newpage





\end{document}